\documentclass[journal,10pt, twocolumn, final]{IEEEtran}
\pdfoutput=1

\usepackage[]{graphicx}    
\newcommand{\ignore}[1]{}  
\usepackage[dvipsnames, table]{xcolor}
\usepackage{hhline}
\usepackage[labelfont=bf, justification=centering]{caption}
\usepackage{float} 
\usepackage{multirow}
\usepackage{float}
\usepackage{url}
\graphicspath{{Figures/}}
\usepackage{amsmath}
\usepackage{soul}
\usepackage{comment}
\usepackage{balance}
\usepackage{gensymb}
\usepackage{cite}
\newcolumntype{M}[1]{>{\centering\arraybackslash}m{#1}}
 \usepackage[compact]{titlesec}
\titlespacing{\section}{0pt}{1ex}{0.5ex}
\titlespacing{\subsection}{0pt}{1ex}{0ex}
\titlespacing{\subsubsection}{0pt}{0.5ex}{0ex}
\usepackage{paralist}
\usepackage[left=0.625in,
            right=0.625in,
            top=0.75in,
            bottom=0.9in]{geometry}
\usepackage[acronyms,nonumberlist,nopostdot,nomain,nogroupskip]{glossaries}
\newacronym{mmwave}{mmWave}{millimeter wave}
\newacronym{phy}{PHY}{physical layer}
\newacronym{mac}{MAC}{medium access control}
\newacronym{uav}{UAV}{unmanned autonomous vehicle}
\newacronym{em}{EM}{electromagnetic}
\newacronym{iot}{IoT}{Internet of Things}
\newacronym{ml}{ML}{machine learning}
\newacronym{drl}{DRL}{deep reinforcement learning}
\newacronym{urc}{URC}{ultra-reliable computing}
\newacronym{urllc}{URLLC}{ultra-reliable low-latency communication}
\newacronym{miso}{MISO}{multiple-input single-output}
\newacronym{mimo}{MIMO}{multiple-input multiple-output}
\newacronym{mu}{MU}{multi-user}
\newacronym{rfid}{RFID}{Radio Frequency Identification}
\newacronym{rfp}{RFP}{radio fingerprinting}
\newacronym{sdr}{SDR}{software-defined radio}
\newacronym{mas}{MAS}{Mobile Autonomous System}
\newacronym{rl}{RL}{reinforcement learning}
\newacronym{los}{LoS}{line of sight}
\newacronym{dnn}{DNN}{deep neural network}
\newacronym{fpga}{FPGA}{field-programmable gate array}
\newacronym{cv}{CV}{computer vision}
\newacronym{mcs}{MCS}{modulation and coding scheme}
\newacronym{soc}{SoC}{system-on-chip}
\newacronym{mumimo}{MU-MIMO}{\gls{mu}-\gls{mimo}}
\newacronym{dsp}{DSP}{digital signal processing}
\newacronym{snr}{SNR}{signal-to-noise ratio}
\newacronym{csi}{CSI}{channel state information}
\newacronym{svd}{SVD}{singular value decomposition}
\newacronym{ism}{ISM}{industrial, scientific and medical}
\newacronym{dsa}{DSA}{dynamic spectrum access}
\newacronym{cnn}{CNN}{convolutional neural network}
\newacronym{rfsoc}{RFSoC}{RF System on Chip}
\newacronym{ntp}{NTP}{network time protocol}
\newacronym{ptp}{PTP}{precise time protocol}
\newacronym{nu}{NU}{Northeastern University}
\newacronym{tamu}{TAMU}{Texas A\&M University}
\newacronym{ncsu}{NCSU}{North Carolina State University}
\newacronym{uta}{UTA}{University of Texas Austin}
\newacronym{fiu}{FIU}{Florida International University}
\newacronym{uo}{OU}{University of Oklahoma}
\newacronym{gt}{GT}{Georgia Tech}
\newacronym{ucb}{UCB}{University of California Berkeley}
\newacronym{ucsb}{UCSB}{University of California Santa Barbara}
\newacronym{ttu}{TTU}{Texas Tech University}
\newacronym{uh}{UH}{University of Hawaii}
\newacronym{eess}{EESS}{Earth exploration satellite services}
\newacronym{nsf}{NSF}{National Science Foundation}
\newacronym{ntia}{NTIA}{National Telecommunications and Information Administration}
\newacronym{rfc}{RFC}{Request for Comments}
\newacronym{dod}{DoD}{Department of Defense}
\newacronym{frp}{FRP}{foundational research principle}
\newacronym{fec}{FEC}{forward error and erasure correction coding}
\newacronym{cct}{CCT}{technological cross-cutting theme}
\newacronym{wiot}{WIoT}{Institute for the Wireless Internet of Things}
\newacronym{ewd}{EWD}{education and workforce development}
\newacronym{pawr}{PAWR}{Platforms for Advanced Wireless Research}
\newacronym{ppo}{PPO}{\gls{pawr} Project Office}
\newacronym{iq}{IQ}{in-phase and quadrature}
\newacronym{if}{IF}{intermediate}
\newacronym{pl}{PL}{physical logic}
\newacronym{lna}{LNA}{low-noise amplifier}
\newacronym{rf}{RF}{radio frequency}
\newacronym{wpan}{WPAN}{wireless personal area networks}
\newacronym{wlan}{WLAN}{wireless local area networks}
\newacronym{wan}{WAN}{Wide Area Networks}
\newacronym{6g}{6G}{sixth generation}
\newacronym{vr}{VR}{virtual reality}
\newacronym{ar}{AR}{augmented reality}
\newacronym{src}{SRC}{Semiconductor Research Corporation}
\newacronym{darpa}{DARPA}{Defense Advanced Research Projects Agency}
\newacronym{adc}{ADC}{analog to digital converter}
\newacronym{dac}{DAC}{digital to analog converter}
\newacronym{gsps}{GSps}{Gigasamples-per-second}
\newacronym{awg}{AWG}{arbitrary waveform generator}
\newacronym{dso}{DSO}{digital storage oscilloscope}
\newacronym{nlos}{NLoS}{non line of sight}
\newacronym{thz}{THz}{terahertz}
\newacronym{si}{Si}{silicon}
\newacronym{soi}{SoI}{Silicon-on-Insulator}
\newacronym{sige}{SiGe}{Silicon-Germanium}
\newacronym{inp}{InP}{Indium Phosphide}
\newacronym{gan}{GaN}{Gallium Nitride}
\newacronym{gaas}{GaAs}{Gallium Arsenide}
\newacronym{jpl}{JPL}{Jet Propulsion Laboratory}
\newacronym{ic}{IC}{integrated circuit}
\newacronym{hbt}{HBT}{heterojunction bipolar transistor}
\newacronym{hemt}{HEMT}{high-electron mobility transistor}
\newacronym{pa}{PA}{power amplifier}
\newacronym{hdl}{HDL}{hardware description language}
\newacronym{fft}{FFT}{fast Fourier transform}
\newacronym{css}{CSS}{chirp spread spectrum}
\newacronym{dsss}{DSSS}{direct-sequence spread spectrum}
\newacronym{rssi}{RSSI}{received signal strength indicator}
\newacronym{bs}{BS}{base station}
\newacronym{ue}{UE}{user equipment}
\newacronym{nrdz}{NRDZ}{National Radio Dynamic Zone}
\newacronym{ofdm}{OFDM}{Orthogonal Frequency Division Multiplexing}
\newacronym{trl}{TRL}{technology readiness level}
\newacronym{ldgm}{LDGM}{low-density generator matrix}
\newacronym{ldpc}{LDPC}{low-density parity-check}
\newacronym{lo}{LO}{local oscillator}
\newacronym{isec}{ISEC}{Interdisciplinary Science and Engineering Complex}

\newacronym{osa}{OSA}{OpenAirInterface Software Alliance}
\newacronym{casper}{CASPER}{Collaboration for Astronomy Signal Processing and Electronics Research}
\newacronym{qos}{QoS}{Quality of Service}
\newacronym{oran}{O-RAN}{Open Radio Access Network}
\newacronym{ran}{RAN}{Radio Access Network}

\newacronym{ric}{RIC}{RAN Intelligent Controller}
\newacronym{cbrs}{CBRS}{Citizens Broadband Radio Service}
\newacronym{gaa}{GAA}{General Authorized Access}
\newacronym{pal}{PAL}{Priority Access Licensee}
\newacronym{fcc}{FCC}{Federal Communications Commission}
\newacronym{sas}{SAS}{spectrum access system}
\newacronym{ai}{AI}{artificial intelligence}
\newacronym{irs}{IRS}{intelligent reflecting surface}

\newacronym{ser}{SER}{symbol error rate}
\newacronym{rfic}{RFIC}{radio frequency integrated circuit}
\newacronym{rfi}{RFI}{\gls{rf} Interference}
\newacronym{aml}{AML}{adversarial machine learning}
\newacronym{sdn}{SDN}{software-defined networking}
\newacronym{star}{STAR}{Simultaneous Transmit and Receive}
\newacronym{sinr}{SINR}{signal-to-interference-noise ratio}
\newacronym{vlba}{VLBA}{Very Long Baseline Array}
\newacronym{ngvla}{ngVLA}{Next Generation Very Large Array}
\newacronym{nrao}{NRAO}{National Radio Astronomy Observatory}
\newacronym{fso}{FSO}{Free Space Optics}
\newacronym{ngso}{NGSO}{non-geostationary orbit}
\newacronym{vsd}{VSD}{Value-Sensitive Design}
\newacronym{sensr}{SENSR}{Spectrum Efficient National Surveillance Radar}
\newacronym{gbps}{Gbps}{Gigabit-per-second}
\newacronym{tbps}{Tbps}{Terabit-per-second}
\newacronym{nas}{NAS}{Network Attached Storage}
\newacronym{5gb}{5GB}{5G-and-beyond}
\newacronym{osi}{OSI}{Open Systems Interconnection}
\newacronym{onr}{ONR}{Office of Naval Research}
\newacronym{afosr}{AFOSR}{Air Force Office of Scientific Research}
\newacronym{afrl}{AFRL}{Air Force Research Laboratory}
\newacronym{arl}{ARL}{Army Research Laboratory}
\newacronym{bdss}{BDSS}{broadband directional spectrum sensor}
\newacronym{circ}{CIRC}{Community Infrastructure for Research in Computer and Information Science and Engineering}

\newacronym{aoa}{AoA}{Angle of Arrival}
\newacronym{noe}{NOE}{NRDZ Orchestration Engine}

\newacronym{mchem}{MCHEM}{Massive Channel Emulator}
\newacronym{afc}{AFC}{Automated Frequency Coordination}
\newacronym{esc}{ESC}{Environmental Sensing Capability}

\newacronym[firstplural=Devices Under Test (DUTs)]{dut}{DUT}{Device Under Test}

\newacronym{kpi}{KPI}{Key Performance Indicator}
\newacronym{dei}{DEI}{diversity, equity and inclusion}

\newacronym{itu}{ITU}{International Telecommunication Union}
\newacronym[firstplural=Notices of Inquiry (NOIs)]{noi}{NOI}{Notice of Inquiry}
\newacronym{wp}{WP}{Working Party}
\newacronym{drs}{DRS}{Digital Repository Service}
\newacronym{ms}{M.S.}{Master of Science}

\newacronym{lxc}{LXC}{Linux Container}
\newacronym{vpn}{VPN}{Virtual Private Network}
\newacronym{ldap}{LDAP}{Lightweight Directory Access Protocol}
\newacronym{ntn}{NTN}{non-terrestrial network}
\newacronym{xr}{XR}{eXtended Reality}
\newacronym{leo}{LEO}{low-Earth orbit}
\newacronym{hap}{HAP}{high-altitude platform}
\newacronym{isl}{ISL}{inter-satellite link}
\newacronym{ul}{UL}{uplink}
\newacronym{dl}{DL}{downlink}
\newacronym{vsat}{VSAT}{Very Small Aperture Terminal}
\newacronym{swap}{SWaP}{size, weight, and power}
\newacronym{cgr}{CGR}{Contact Graph Routing}
\newacronym{dtn}{DTN}{Delay Tolerant Network}
\newacronym{apd}{APD}{avalanche photo detector}
\newacronym{arq}{ARQ}{Automatic Repeat Request}
\newacronym{csli}{CSLI}{CubeSat Launch Initiative}
\newacronym{csdw}{CSDW}{CubeSat Developer Workshop}
\newacronym{ssc}{SSC}{Small Satellite Conference}
\newacronym{icc}{ICC}{International Conference on Communications}
\newacronym{api}{API}{Application Programming Interface}
\newacronym{3gpp}{3GPP}{3rd Generation Partnership Project}
\newacronym{ttc}{TT\&C}{Telemetry, Tracking and Control}
\newacronym{adcs}{ADCS}{attitude determination and control system}
\newacronym{pms}{PMS}{Power Management System}
\newacronym{ber}{BER}{Bit Error Rate}
\newacronym{cs}{CS}{Computer Science}
\newacronym{cise}{CISE}{Computer and Information Science and Engineering}
\newacronym{hpa}{HPA}{High Power Amplifier}
\newacronym{seu}{Single Event Upsets}{SEU}
\newacronym{ots}{OTS}{off-the-shelf}
\newacronym{msu}{MSU}{Morehead State University}
\newacronym{ppdr}{PPDR}{Public Protection and Disaster Relief}
\newacronym{sigmf}{SigMF}{Signal Metadata Format}
\newacronym{sscm}{SSC}{Space Science Center}
\newacronym{unp}{UNP}{University Nanosatellite Porgram}
\newacronym{unlab}{UNLab}{Ultrabroadband Nanonetworking Laboratory}
\newacronym{coe}{COE}{College of Engineering}
\newacronym{aeronu}{AeroNU}{AerospaceNU}
\newacronym{dsn}{DSN}{Deep Space Network}
\newacronym{click}{CLICK}{CubeSat Laser Infrared Cross-link}
\newacronym{tbird}{TBIRD}{Terabyte Infrared Delivery}
\newacronym{mc}{MC}{Mission Concept}
\newacronym{stem}{STEM}{Science, Technology Engineering, and Mathematics}
\newacronym{knrc}{KNRC}{Kostas Nanomanufacturing Research Center}
\newacronym{sdl}{SDL}{Space Dynamics Laboratory}
\newacronym{obc}{OBC}{On Board Computer}
\newacronym{gs}{GS}{Ground Station}
\newacronym{cots}{COTS}{commercial off-the-shelf}
\newacronym{ussf}{USSF}{U.S. Space Force}
\newacronym{usaf}{USAF}{U.S. Air Force}
\newacronym{iss}{ISS}{International Space Station}
\newacronym{bpsk}{BPSK}{Binary Phase Sfhit Keying}
\newacronym{conops}{ConOps}{Concept of Operations}
\newacronym{lsp}{LSP}{Launch Service Provider}
\newacronym{eps}{EPS}{Electric Power System}
\newacronym{cdh}{CD\&H}{Command and Data Handling}

\newacronym{scr}{SCR}{System Concept Review}
\newacronym{srr}{SRR}{System Requirement Review}
\newacronym{rvm}{RVM}{Requirement Verification Matrix}
\newacronym{pmr}{PMR}{Program Management Review}
\newacronym{pdr}{PDR}{Preliminary Design Review}
\newacronym{cdr}{CDR}{Critical Design Review}
\newacronym{fsr}{FSR}{Flight Selection Review}
\newacronym{pn}{PN}{Pseudorandom Noise}
\newacronym{mls}{MLS}{Microwave Limb Sounder}
\newacronym{pdp}{PDP}{Power Delay Profile}
\newacronym{udp}{UDP}{User Datagram Protocol}
\newacronym{cfo}{CFO}{Carrier Frequency Offset}
\newacronym{eof}{EoF}{Enf of Frame}
\newacronym{sll}{SLL}{Side Lobe Level}
\newacronym{vhf}{VHF}{Very High Frequency}
\newacronym{uhf}{UHF}{Ultra High Frequency}
\newacronym{satnogs}{SATNOGS}{Satellite Networked Open Ground Station}
\newacronym{tle}{TLE}{Two-Line Element}
\newacronym{qam}{QAM}{Quadrature-Amplitude Modulation}
\newacronym{isara}{ISARA}{Integrated Solar Array and Reflectarray Antenna}
\newacronym{hitran}{HITRAN}{HIgh resolution TRANsmission molecular absorption}
\newacronym{eirp}{EIRP}{effective isotropic radiated power}

\begin{document}
\bstctlcite{IEEEexample:BSTcontrol}

\title{Sub Terahertz LEO Satellite Communication: Vision,~Opportunities,~and~Challenges\\toward the First Prototype in Space}

\author{Sergi Aliaga, Vitaly Petrov, Andrew Benincasa, Albert Diez, Ali J. Alqaraghuli, Jose V. Siles, Ken R. Duffy, Marc~Sanchez~Net, Tommaso Melodia, and Josep M. Jornet

\thanks{S. Aliaga, A. Benincasa, A. Diez, K. Duffy, T. Melodia, and J. M. Jornet are with Northeastern University, Boston, MA, USA. V. Petrov is with KTH Royal Institute of Technology, Stockholm, Sweden. A. J. ALqaraghuli is with Teralink Technologies, Cambridge, MA, USA. J. V. Siles, and M. Sanchez Net are with NASA Jet Propulsion Laboratory at CalTech, Pasadena, CA, USA. Part of the research was carried out at the Jet Propulsion Laboratory, California Institute of Technology, under a contract with the National Aeronautics and Space Administration (80NM0018D0004). \emph{Corresponding author: Sergi Aliaga}}%
\vspace{-10mm}
}

\maketitle

\begin{abstract}
The landscape of sub-terahertz (sub-THz, 100\,GHz--300\,GHz) wireless technology evolved drastically over the last two decades -- from only a few niche use cases in sensing and ultra-short-range communications in early 2000s toward operational multi-kilometer range 100\,GBbit/s+ wireless backhaul links demonstrated recently. Building on this momentum, this article explores the feasibility of extending sub‑THz communications to 100‑km‑scale satellite links. We first assess the technological readiness of emerging sub‑THz hardware and signal‑processing techniques, highlighting their potential to support long‑range operation in low‑Earth‑orbit (LEO) systems. We then outline the unique role that sub‑THz links can play as a complementary solution to existing millimeter‑wave and optical (``laser'') satellite technologies, offering additional capacity, improved resilience, and new architectural flexibility. We further discuss open research and engineering challenges toward implementing such sub-THz satellite communication systems in practice. We finally outline the key state-of-the-art solutions and the roadmap of \emph{TeraLink}, an ongoing international R\&D project aiming to build and launch, through an approved NASA CSLI space mission,  \emph{the first hardware prototype of sub-THz LEO satellite communications in space}.\vspace{-3mm}
\end{abstract}

\section{Introduction}
\label{sec:intro}
The technology landscape of wireless systems operating in the terahertz band (THz, 300\,GHz--3\,THz) and, especially, sub-terahertz band (sub-THz, 100\,GHz--300\,GHz, sometimes referred to as ``higher millimeter wave (mmWave) bands'') changed dramatically over the recent decades. Following a few visionary works from early 1980s, the exploration of ``sub-terahertz wireless communications'' began globally in the late 1990s and early 2000s. At that time, many of the studies concluded that the combination of: (i)~inherently high spreading loss at sub-terahertz frequencies; (ii)~low efficiency of sub-THz electronics sometimes also referred to as ``the THz gap''; and (iii)~the desire to unleash the full potential of wide sub-terahertz spectrum, leading to theoretically higher data rates, but also spreading already limited transmit power over wider bands, would very likely limit the maximum communication range between the nodes equipped with THz and sub-THz radios. (Sub-)THz wireless communication was then explored by the community primarily as a solution for future ultra-short-range wireless links (e.g., between different cores of a multi-core CPU) and so-called ``nanonetworks''~\cite{akyildiz2022terahertz}.

However, the progress in this field over the recent 30 years substantially challenges this original short-range-only assumption. First, the recent technology developments in the field of higher-power and lower-noise sub-THz electronics already allow to claim the so-called ``THz Gap'' partially closed (at least, in the sub-THz part of the spectrum)~\cite{cooper2025power}. For instance, sub-THz transmitters with 30+\,dBm of \gls{eirp} have been demonstrated, which already, e.g., notably exceeds the EU EIRP regulations for Wi-Fi 2.4\,GHz band (\gls{eirp} of 100\,mW = 20\,dBm)~\cite{gonzalez2024dband}. Second, the appearance of novel compact, high-efficiency, and high-directivity sub-THz antennas allowed to partially address the spreading loss problem and improve coverage, as the combination of high-gain transmitter and high-gain receiver antennas boosts the overall link budget by as high as 80~dB~\cite{maestrojuan2018reflector}. Third, the 30+ years of innovation  in signal processing techniques allows notably more robust decoding of a wideband sub-THz signal even in low signal-to-noise ratio regimes~\cite{takashima2024fundamental}.

The combination of all these has allowed sub-THz communication links to go far beyond some earlier assumptions of cm-scale, meter-scale, or even tens/hundreds of meters of coverage scale (for, e.g., next-generation WLANs). Several research teams, from university labs, through industry players, to defense R\&D centers, such as the U.S. AFRL, have demonstrated \textbf{multi-km-long} high-rate wireless links operating in different parts of the sub-THz spectrum~\cite{sen2023multi}. Such verified-in-hardware results confirm the feasibility of multi-kilometer-long sub-THz wireless links today, with state-of-the-art technology. Meanwhile, even better performance and lower-cost systems are expected to appear in the future. Hence, a \emph{terrestrial} multi-km link between two towers using sub-THz radio is already proven to be feasible.

\emph{In this article, we make one step further and advocate for the utilization of sub-THz radio not just for 2--10\,km terrestrial wireless links, but for 100\,km+ satellite communication links, particularly for high-rate connectivity with and between low-Earth-orbit (LEO) satellites}. While hardware prototypes have already validated most shorter-range sub-THz communication use cases, satellite sub-THz connectivity remains unexplored despite its significant potential. Here, our \emph{TeraLink mission} -- an ambitious ongoing international collaboration described in this article -- seeks to address this gap by providing the first practical demonstration of sub-THz communications in the space environment.

The rest of this article and its contributions are structured as follows. We start with a tutorial-in-nature analysis of the advantages and limitations of sub-THz wireless communications for satellite connectivity, compared to already deployed mmWave and optical satellite systems. In particular, we present arguments that there is a gap to be filled by such new radios, combining the strengths of mmWave and optical communications. We then describe the key design aspects of TeraLink and examine the research opportunities to be enabled by this platform, outlining the experimental campaign planned for validation. We finally outline and discuss key future research directions to explore toward realizing practical large-scale, high-performance sub-THz satellite networks.

\section{High-Speed Satellite Technology Landscape: From mmWave and optical to THz Frontiers}
\label{sec:ThzSpaceComms}
Tremendous advancements in \gls{mmwave} and optical satellite communication have occurred in recent years, partially due to the development of small satellite platforms known as CubeSats. CubeSats follow standardized sizes (1U, 2U, etc.), where ``U" denotes a $10\times10\times10~\text{cm}^3$ unit, and have proven invaluable as cost-effective, scalable, and easy-to-deploy platforms for various LEO space missions.

\subsection{Millimeter Wave CubeSats}
A key milestone in mmWave satellite experimentation is the \gls{isara}, a 3U satellite launched by NASA in 2017~\cite{ISARA}. It demonstrated a deployable Ka-band ($26$~GHz) reflectarray antenna integrated with a solar panel, achieving \gls{dl} speeds of $100$~Mbps. \gls{isara} validated that compact, high-gain antennas can support high-rate \gls{mmwave} CubeSat satellite links, even within constrained size and power budgets.

Another notable mission is KIPP, the first satellite deployed by Kepler Communications in 2018 to deliver high-throughput Ku-band ($12$\,GHz--$18$\,GHz) services from low Earth orbit~\cite{laverty2020kepler}. Built around a custom software-defined radio and a high-gain antenna, KIPP demonstrated \gls{ul} speeds up to $150$~Mbps, with a theoretical limit of $300$~Mbps. Operating as part of Kepler’s broader plan to establish a space-based Internet, KIPP proved that CubeSats can deliver broadband services to underserved regions and validated the use of dynamic, reprogrammable radios in space-based mmWave systems.

However, unlike optical or sub-THz systems discussed further, the mmWave system bandwidth may not exceed few (tens of) GHz. This is sufficient for many practical \gls{ul} and \gls{dl} scenarios, but not for all the use cases (e.g., not for high-rate \glspl{isl} multiplexing tens/hundreds of parallel user transmissions). Hence, the satellite community recently started exploring even higher-frequency communication systems in the optical and (sub-)THz frequency bands.

\begin{figure*}[h!]
    \centering
    \includegraphics[width=\linewidth]{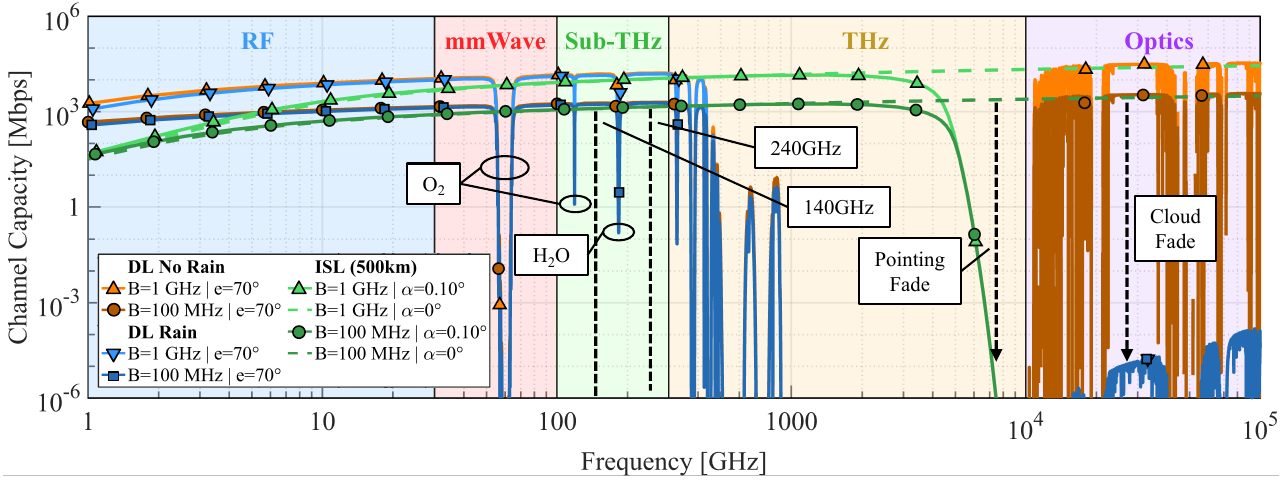}
    \caption{Channel capacity comparison across frequency bands for different satellite links, accounting for spreading loss, atmospheric absorption, rain attenuation, and pointing loss.}
    \label{fig:channelCapacity}
    \vspace{-6mm}
\end{figure*}

\subsection{Free-Space Optical CubeSats}
In the \gls{fso} domain, a 3U twin CubeSat platform~\cite{Cahoy2019Cubesat} is targeting \gls{isl} and \gls{ul}/\gls{dl} tests at optical frequencies. In this case, the limiting factor is the $1$~GHz bandwidth of the \gls{apd}, resulting in maximum data rates of $17.7$~Mbps for \glspl{isl} and $10$~Mbps for \gls{dl}. Meanwhile, a 6U CubeSat platform proposed in~\cite{Schieler2022tbird} was aimed at testing novel \gls{arq} protocols over optical links, targeting burst \gls{dl} rates of up to $200$~Gbps, but requiring complex beam alignment mechanisms and protocols due to the tight laser beam width.

Furthermore, while the use of \gls{fso} for \gls{isl} connectivity seems promising~\cite{alimi2024revolutionizing}, the future of \gls{ul}/\gls{dl} \gls{fso} remains questionable. Beyond experimental CubeSats, large-scale commercial mega-constellations, like Starlink or Amazon Leo, increasingly rely on hybrid FSO and RF/mmWave architectures to balance high-capacity inter-satellite routing with environmental robustness. Even within these resource-rich networks, optical ground-to-space links remain highly vulnerable to atmospheric degradation~\cite{Aliaga2025analysis}, reinforcing the critical need for weather-resilient alternatives.

\subsection{The opportunity for THz satellite communication}
THz communication is emerging as a compelling possibility for broadband space communications, offering advantages over its \gls{mmwave} and \gls{fso} counterparts. These include greater robustness to atmospheric conditions (in \gls{ul}/\gls{dl}), relaxed pointing requirements (in \gls{isl}), and a compact hardware footprint.

It is worth noting that the potential of THz technology in space was recognized well before the current surge in satellite research. Specifically, due to its relatively better weather resilience compared to \gls{fso}, the utility of the THz band has long been acknowledged by the radio astronomy community -- even prior to the launch of the \gls{mls} instrument aboard NASA’s Aura satellite (2004). 

All solid-state THz sources and receivers have demonstrated tremendous progress over the past three decades and represent a mature technology that has flown multiple times in space. Notable additional missions include the HIFI instrument on Herschel Space Observatory~(2009), and the MIRO instrument on the Rosetta Orbiter (2004). Moreover, significant advances in power generation over the last 15 years have enabled THz imaging radars for Earth Science applications -- such as VIPR and NASA's upcoming CloudCube mission\cite{cloudcube}, which builds on RainCube's Ka-band radar for precipitation sensing and extends passive observation into the sub-THz range (239 GHz) -- as well as array receivers for astrophysics missions including NASA's STO-2 and ASTHROS. 

These advances, particularly in THz power generation~\cite{Siles2018New}, now make it possible for the first time to develop sub-THz satellite communication links. Nonetheless, none of the discussed prototypes had an active THz transmitter designed for high-rate communication -- a gap that TeraLink aims to address.

\subsection{Strengths and Weaknesses: a Comparison}
As an overview, Fig.\ref{fig:channelCapacity} presents the channel capacity as a function of carrier frequency for two distinct scenarios:  \gls{dl} between a \gls{gs} and a CubeSat in \gls{leo} ($400$~km altitude), and an \gls{isl} between two CubeSats separated by $500$~km. For the ground-to-space link, the assessment includes both clear-sky and rainy conditions, shown in orange and blue, respectively, and assumes a $70^{\circ}$ elevation angle. Atmospheric losses, including both molecular absorption and scattering, are computed using ITU Recommendation ITU-R P.676-12~\cite{ITU676}, as well as the absorption data from the \gls{hitran} line catalog. Each environmental condition includes two curves corresponding to different channel bandwidths. In the \gls{isl} scenario, while the link is not affected by atmospheric attenuation, additional capacity curves are included to account for potential pointing error ($\alpha$) between the spacecraft, which introduce additional losses. The results are obtained assuming a state-of-the-art sub-THz communication system. Specifically, the link budget considers a transmit power of $25$~dBm, a CubeSat antenna with a 1U aperture ($10\times10\text{cm}^2$), a $1$~m diameter \gls{gs} dish with $60$~\% efficiency, a receiver noise figure of $7$~dB, and an antenna noise temperature of $300$~K.

Focusing on the \gls{ul} and \gls{dl} scenarios, the capacity curves reveal several key trends across the frequency spectrum. In the RF band, channel capacity remains relatively unaffected by atmospheric absorption, as expected. However, as the frequency increases into the \gls{mmwave} regime, a pronounced absorption notch appears around $60$~GHz due to the resonance with oxygen molecules. Beyond this point, multiple absorption peaks emerge in the sub-THz range, growing denser and more intense as the frequency approaches the mid-THz region. In this range, atmospheric gases absorb energy so strongly that sustaining satellite communication becomes extremely challenging. This effect can be advantageous for (Sub)-THz \glspl{isl} as robustness to ground eavesdropping is intrinsically associated with the choice of a carrier frequency that presents large absorption by atmospheric gases (i.e. 183GHz). Beyond this absorption-heavy band, channel capacity increases again in the optical spectrum, although additional absorption features remain present.

One of the most significant insights from the figure is the distinct sensitivity of different frequency bands to weather conditions. In clear-sky conditions, optical frequencies offer the highest channel capacity among all bands. However, their performance degrades drastically in the presence of clouds and rain, effectively rendering the link unusable. In contrast, the sub-THz and lower frequency bands exhibit minimal difference between clear and rainy conditions, highlighting their robustness to atmospheric impairments.

Shifting focus to the \gls{isl}, the capacity curves are much smoother, as atmospheric absorption is absent and only free-space path loss and pointing error ($\alpha$) are present. When no pointing error is present, the \gls{isl} achieves its maximum capacity in the optical range, owing to the highly directional nature of optical beams and the favorable aperture-to-wavelength ratio. However, for the same reason, even sub-degree misalignments introduce significant pointing losses, which rapidly degrade the capacity beyond the upper-THz range. This fragility explains the need for complex and bulky pointing payloads to maintain precise alignment of optical transceivers in space, which in turn increases mission cost and places a significant burden on the spacecraft’s \gls{swap} budget.

Analyzing the effect of bandwidth, it becomes evident that in the RF and \gls{mmwave} regimes, the link operates in a power-limited regime due to the small satellite aperture and relatively large wavelengths. In contrast, as the frequency increases into the sub-THz domain, the link transitions into a bandwidth-limited regime, where increasing the bandwidth leads to substantial gains in capacity. This shift is driven by the improved directivity and reduced beam divergence at higher frequencies.

Taken together, the results clearly demonstrate the advantages of THz -- and more specifically, sub-THz -- communications for space applications. \emph{This frequency range offers significantly higher channel capacity than conventional RF and mmWave bands, while maintaining greater resilience to atmospheric losses and pointing errors than optical systems}. As such, the sub-THz band represents a promising middle ground for future high-throughput satellite links. Of particular interest are two low-absorption windows in this range, centered at $140$~GHz and $240$~GHz, each offering bandwidths of at least $10$~GHz, and achieving \glspl{snr} around $16.4$~dB and $19.8$~dB, respectively. Notably, the second of these singular sub-THz windows i.e., $240$~GHz, was selected for TeraLink, as described next.

\section{TeraLink: An experimental prototype for THz Communications in space}\label{sec:TeraLink}
Building on the outlined advantages, we introduce TeraLink: the first CubeSat mission designed to demonstrate and characterize sub-THz communications in space, establishing proof-of-concept for multi-gigabit satellite links above $200$~GHz. The mission pursues a clear goal:
\begin{center}
\textit{Establish sub-THz access links (\gls{ul} and \gls{dl}) to and from LEO to enable high-speed communication with small spacecraft.}
\end{center}

This goal is decomposed into two technical thrusts, each with measurable success criteria. First, TeraLink will demonstrate link closure by detecting an unmodulated sub-THz carrier with SNR$>0$~dB, validating the readiness of sub-THz hardware for space. Second, it will demonstrate high-rate communication by transmitting at least $100$~Mbps with an uncoded \gls{ber} below $10^{-4}$. Both of these thrusts prioritize the downlink, recognizing the typical asymmetry in data traffic compared to the satellite uplink.

\begin{figure}[h]
    \centering
    \includegraphics[width=\linewidth]{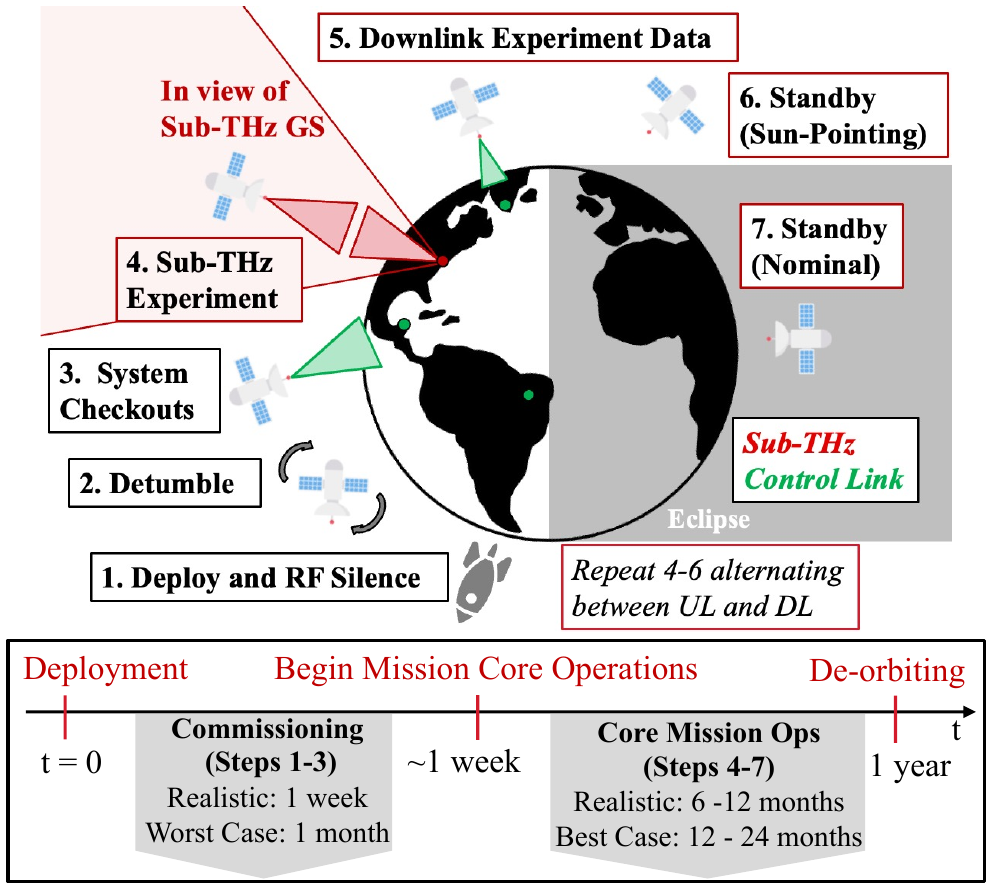}
    \caption{TeraLink Mission Concept of Operations.}
    \label{fig:Conops}
    \vspace{-3mm}
\end{figure}

The mission, implemented as a 6U CubeSat, follows a two-phase operational concept (Fig.~\ref{fig:Conops}). After a brief commissioning phase ($\approx$$1$ week), the satellite enters Core Mission Operations, where it cyclically executes sub-THz experiments during ground station passes. Each pass alternates between \gls{dl} and \gls{ul} experiments based on ground commands. To maximize experimentation opportunities, experiments may also proceed during eclipse periods when the satellite is in view of the sub-THz ground station located in Boston, USA. The Core Mission Operations phase is expected to span 6–12 months, with an upper bound of 24 months, concluding with natural orbital decay in compliance with NASA's \gls{csli} five-year orbital lifetime limit.

\begin{figure}[h]
    \centering
    \vspace{-3mm}
    \includegraphics[width=\linewidth]{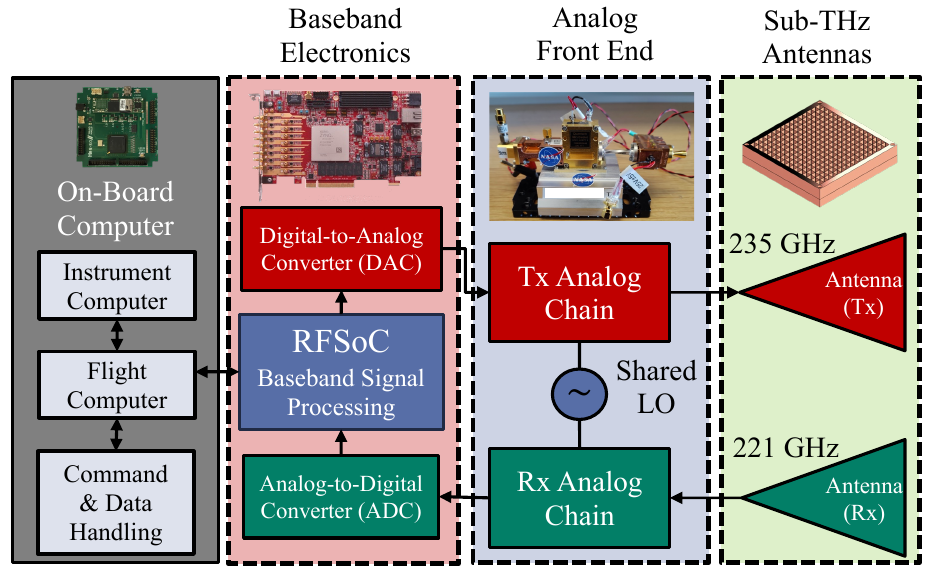}
    \caption{Sub-THz SDR Architecture for the TeraLink satellite.}
    \label{fig:payload}
    \vspace{-3mm}
\end{figure}

The core payload of the satellite is a sub-THz \gls{sdr} operating at $209$–$240$~GHz, composed of three main blocks (Fig.~\ref{fig:payload}): custom-made compact horn antenna arrays, a multiplier-based analog front-end transceiver, and a high-speed digital signal processing engine. The radio uses a variable local oscillator to operate at $217$–$226$ GHz for \gls{ul} and $232$–$240$ GHz for \gls{dl} -- frequency bands allocated by the FCC for space communication~\cite{sen2023multi} that align with the low-absorption atmospheric windows identified in Section~\ref{sec:ThzSpaceComms}.

\subsection{Horn Antenna Array}
\begin{figure}[h]
    \centering
    \includegraphics[width=\linewidth]{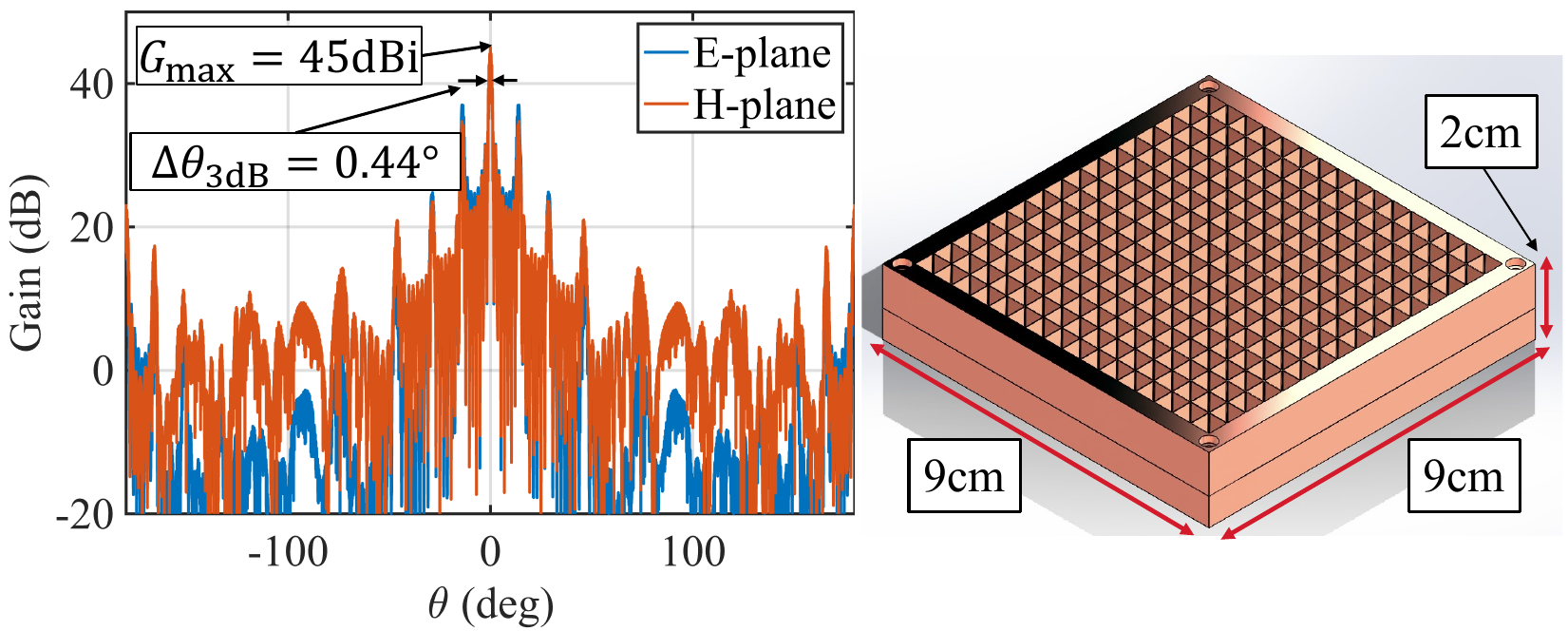}
    \caption{Horn Antenna Array design for the TeraLink satellite.}
    \label{fig:Antenna}
\end{figure}

A pair of custom-made horn antenna arrays provide $45$~dBi gain while fitting within a 6U CubeSat form factor. Each $16\times16$ element array occupies a 1U face ($10$~cm$\times 10$~cm), with a planar structure measuring $9$cm$\times9$~cm (Fig.~\ref{fig:Antenna}). This compact design exemplifies a key advantage of sub-THz frequencies: antenna size scales with wavelength, enabling dramatic miniaturization. An equivalent X-band antenna would span over $3$m in its longest dimension -- 20 times larger -- making integration impractical for small satellites. The design is currently being finalized before prototype manufacturing and testing.

\subsection{Sub-THz Front End}
The sub-THz analog front end comprises a super-heterodyne transceiver with a shared local oscillator, employing NASA JPL-pioneered Schottky-diode technology~\cite{cooper2025power}. The transmitter delivers $315$~mW ($25$~dBm) output power -- among the highest demonstrated at sub-THz frequencies -- while the receiver uses a low-noise balanced mixer. Critically, this configuration offers exceptional temperature stability, eliminating active thermal control systems and enhancing compactness. The design has been validated in terrestrial multi-kilometer, multi-Gbps demonstrations at $210$–$230$~GHz~\cite{sen2023multi}, with space qualification currently underway.
\begin{figure*}[ht!]
    \centering
    \includegraphics[width=0.85\linewidth]{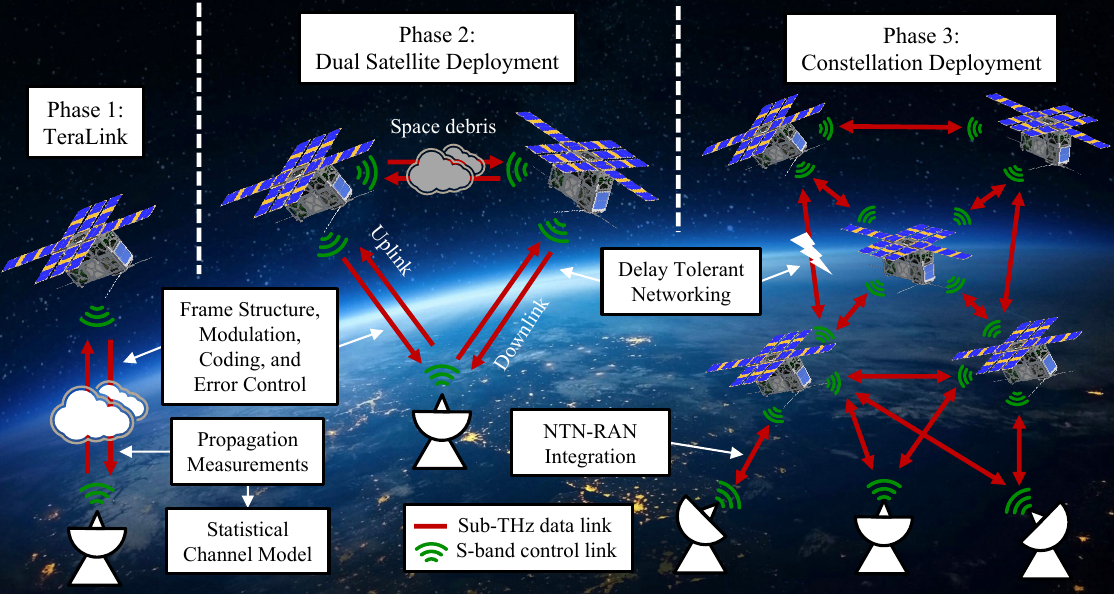}
    \caption{TeraLink development roadmap, from a single-satellite sub-THz link demonstration to a dual-satellite system and, ultimately, a small constellation enabling networking and advanced protocol testing.}
    \label{fig:vision}
    \vspace{-5mm}
\end{figure*}

\subsection{Baseband \gls{dsp} Engine}
The baseband \gls{dsp} engine handles A/D conversion and signal processing using an \gls{rfsoc}, which integrates programmable logic (\gls{fpga}), high-speed data converters (ADCs/DACs), and a processing system (CPU) into a single compact device. The \gls{rfsoc} supports sampling rates exceeding $4$~GSps, enabling the high bit rates necessary for mission objectives, while its flexibility allows real-time reconfiguration for custom \glspl{mcs}.

During transmission, the baseband \gls{dsp} engine manages frame formatting, channel coding, modulation, and pulse shaping. In reception, it records in-phase and quadrature (IQ) samples -- TeraLink's primary data product -- over fixed time windows. The payload supports four phase modulation schemes (BPSK to 16-PSK) and three symbol rates ($8$–$1024$ MBd), yielding multiple configurations exceeding the $100$~Mbps target.

The complete payload -- including baseband radio, analog front-ends, and antennas -- occupies 3–4U within the 6U CubeSat, with specifications summarized in Table~\ref{tab:PayloadSpecs}. Current power budget calculations indicate that the necessary power can be provided in a 6U CubeSat configuration. 
\begin{table}[!h]
    \footnotesize
    \centering
    \begin{tabular}{p{0.45\columnwidth}p{0.42\columnwidth}}
        \hline
        \textbf{Parameter}&\textbf{Value} \\
        \hline
        Carrier Frequency & 221~GHz (UL), 236 (DL) \cite{Siles2018New}\\
        Transmit Power & 25~dBm (315~mW) \cite{Siles2018New}\\
        Noise Figure & 7~dB \cite{Siles2018New}\\
        CubeSat Antenna Gain & 45~dBi (Fig.~\ref{fig:Antenna}) \\
        Maximum Symbol Rate & 1024~GBd \\
        Roll-off factor & 0.3 \\
        Modulation schemes & PSK (M=2, 4, 8, 16) \\
        Sampling Rate & 4.096~GSps \\
        Channel Coding & CA-Polar, LDPC \\
        \hline
    \end{tabular}
    \caption{TeraLink CubeSat payload specifications.}
    \label{tab:PayloadSpecs}
\end{table}

\section{Research Opportunities}
\label{sec:Research}
As summarized above, developing a hardware prototype for an emerging communication technology (sub-THz) is challenging, especially, in space; it requires collective efforts across multiple disciplines (from antenna design through bus development to embedded programming). At the same time, hardware prototypes, such as TeraLink, enhance the credibility of proposed solutions and reveal unanticipated challenges often overlooked in analytical models. Furthermore, there are various research opportunities enabled by TeraLink.

\textbf{a) Propagation Measurements and Channel Sounding:}
By repeatedly transmitting pilot signals to and from \gls{leo}, TeraLink will generate the first statistically meaningful sub-THz satellite channel model (see Fig.~\ref{fig:vision}), capturing path loss, multipath fading, Doppler shifts, and atmospheric absorption effects. This characterization will accelerate hardware development by extending simulator and digital twin capabilities, facilitating ground-based validation of low-TRL solutions, and reducing costly in-orbit testing.

\textbf{b) Physical Layer Optimization:}
TeraLink's \gls{sdr} payload enables transmission of real data (files, images, video), allowing researchers to explore frame structures, modulation schemes, and coding techniques. The platform provides a unique testbed for optimizing frame configurations (header design, cyclic prefix length, equalization), evaluating novel \gls{fec} codes under high data rates, and refining \gls{mcs} adaptation mechanisms for efficient bandwidth utilization.

\textbf{c) Beam Alignment and Adaptive Control:}
TeraLink enables evaluation of alignment strategies: open-loop methods using signal strength and SNR feedback, closed-loop approaches leveraging additional onboard radios for satellite acquisition, and sensing-assisted techniques utilizing star trackers. Beyond alignment, closed-loop systems can dynamically adjust transmit power based on receiver feedback, enhancing energy efficiency while minimizing interference and receiver saturation.

These examples represent a subset of TeraLink's research opportunities. As the platform matures, it will open even broader avenues for advancing sub-THz communications and satellite networking technologies.

\section{What's Next? A roadmap for unlocking THz communication in Space}
\label{sec:Next}
TeraLink marks the beginning of a long-term vision for global high-rate connectivity, disaster response, climate monitoring, and deep space exploration -- all powered by sub-THz communication. Fig.~\ref{fig:vision} illustrates this phased roadmap and emerging research opportunities at each stage:
\vspace{6mm}

\textbf{Phase 1: TeraLink}\\
The first deployment of a sub-THz radio in space, demonstrating \gls{ul} and \gls{dl} from \gls{leo} and enabling the research opportunities outlined in Sec.~\ref{sec:Research}.

\textbf{Phase 2: Dual Satellite Deployment}\\
Deploying paired TeraLink satellites will demonstrate \glspl{isl} at sub-THz frequencies, validating feasibility across all three primary link types (\gls{ul}, \gls{dl}, and \gls{isl}). This phase enables the development of statistical channel models that account for space debris effects, as microscopic particles can cause multipath propagation or interference despite highly directional antennas. Furthermore, the resulting three-node platform (two satellites plus ground station) establishes the first space-based sub-THz network testbed, enabling experimental validation of delay-tolerant networking, software-defined networking, and satellite-integrated \gls{ran} protocols.

\textbf{Phase 3: Constellation Deployment}\\
Once \gls{isl} functionality is verified, this phase will deploy a small constellation of $10$–$20$ satellites, potentially scaling to $100$+ in multi-orbit configurations. This will enable cross-layer security research, including satellite-based protocol evaluation, ground station eavesdropping analysis, angular diversity for anti-jamming, and reflection-based attack mitigation. This phase will transition sub-THz technology from experimental validation to real-world scalability for high-capacity satellite networks.

\section{Conclusions}
\label{sec:conclusions}
{
This paper outlines a forward-looking vision for enabling high-speed wireless communications in space through\hspace{-1mm} (sub-) THz technology. Sub-THz offers a compelling middle ground between mmWave and optical systems by combining high data rates with relaxed pointing requirements and improved atmospheric resilience. As the demand for global broadband access, real-time disaster response, Earth observation, and deep space exploration continues to grow, (sub-)THz bands stand out as a critical enabler of next-generation satellite networks. Realizing this potential requires bridging theoretical advances with in-orbit validation, and emerging platforms like TeraLink play a foundational role in this transition. By demonstrating the feasibility of Gbps-class links above $200$~GHz from \gls{leo} and supporting in-orbit experimentation with real hardware, TeraLink and its subsequent deployment phases unlock new opportunities in channel modeling, adaptive protocols, and scalable architectures. In the long arc of space communication, (sub-)THz technology marks not just the next frequency frontier, but a leap toward truly high-rate, flexible, and reliable space-based connectivity.
}
\balance


\bibliographystyle{IEEEtran}
\bibliography{Bibliography}



\end{document}